\documentclass[letterpaper,12pt]{article}

\usepackage{amssymb,latexsym}

\usepackage{fullpage}
\usepackage{nicefrac}

\usepackage[pdftex]{graphicx} 

\makeatletter \@addtoreset{equation}{section} 
\makeatother

\begin{document} 
\begin{titlepage}
	\thispagestyle{empty} 
	\begin{flushright}
		\hfill{CERN-PH-TH/2009-148}\\
		\hfill{DFPD-09/TH/15}\\
	\end{flushright}
	
	\vspace{35pt} 
	
	\begin{center}
		{ \LARGE{\bf First order flows for $N=2$ extremal \\[3mm]
		black holes and duality invariants }}
		
		\vspace{50pt}
		
		{Anna Ceresole$^a$, Gianguido Dall'Agata$^b$, Sergio Ferrara $^{c,d}$ and Armen Yeranyan$^{d,e}$}
		
		\vspace{15pt}
		
		{\it ${}^a$ INFN, Sezione di Torino $\&$ Dipartimento di Fisica Teorica\\
		Universit\`a di Torino, Via Pietro Giuria 1, 10125 Torino, Italy}
		
		\vspace{15pt}
		
		{\it ${}^b$ Dipartimento di Fisica ``Galileo Galilei'' $\&$ INFN, Sezione di Padova \\
		Universit\`a di Padova, Via Marzolo 8, 35131 Padova, Italy}
		
		\vspace{15pt}
		
		{\it ${}^c$ Department of Physics, CERN Theory Division\\
		CH 1211, Geneva 23, Switzerland}
		
		\vspace{15pt}
		
		{\it ${}^d$ INFN - LNF, \\
		Via Enrico Fermi 40, I-00044 Frascati, Italy}

		\vspace{15pt}
		
		{\it ${}^e$ Department of Physics, Yerevan State University, \\
		Alex Manoogyan St. 1, Yerevan, 0025, Armenia}
		
		\vspace{40pt}
		
		{ABSTRACT} 
	\end{center}
	
	\vspace{10pt}
	
We derive explicitly the superpotential $W$ for the non-BPS branch of $N=2$ extremal black holes in terms of duality invariants of special geometry.
Although this is done for a one-modulus case (the $t^3$ model), the example gives $Z \neq 0$ black holes and captures the basic distinction from previous attempts on the quadratic series (vanishing $C$ tensor) and from the other $Z=0$ cases.
The superpotential $W$ turns out to be a non-polynomial expression (containing radicals) of the basic duality invariant quantities.
These are the same which enter in the quartic invariant $I_4$ for $N=2$ theories based on symmetric spaces.
Using the flow equations generated by $W$, we also provide the analytic general solution for the warp factor and for the scalar field supporting the non-BPS black holes.
\end{titlepage}

\baselineskip 6 mm

\section{Introduction} \label{Intro}

It has long been known (\cite{FKS}-\cite{FGK}) that the properties of the $N=2$ extremal, static, spherically symmetric black holes of Einstein-Maxwell theories coupled to the special K\"ahler geometry of $n$ complex scalar fields $z^i$ are encoded in the effective potential 
\begin{equation}
	V_{BH}=Z{\overline {Z}}+g^{i{\bar\jmath}} D_i Z{\overline {D}}_{\bar\jmath} {\overline {Z}}\, , \label{potential} 
\end{equation}
where $Z(z,\bar z; q,p)$ is the central charge of the $N=2$ supersymmetry algebra (also function of the  charges), $g_{i{\bar\jmath}}=\partial_i\partial_{\bar\jmath} K(z,\bar z)$ is the metric of the scalar $\sigma$-model and $D_i\equiv \partial_i+\frac12 \partial_i K $ is the K\"ahler covariant derivative. 
For supersymmetric configurations, $Z$ plays the role of a superpotential that drives the first order radial flows for the warp factor and scalar fields towards the black hole horizon: 
\begin{equation}
	U'=- e^U |Z|\,,\qquad\  \qquad z'^i=-2 e^Ug^{i{\bar\jmath}}
	\partial_{\bar\jmath} |Z|\, . \label{flows} 
\end{equation}
These flows stop at the critical points of the central charge, $D_i Z = 0$, which are also supersymmetric critical points of the full potential (\ref{potential}), fixing the values of the scalar fields at the horizon in terms of the electric and magnetic charges.
Therefore supersymmetric black hole configurations are solutions of (\ref{flows}) where scalar fields vary from an arbitrary asymptotic value to a universal attractor point at the horizon, depending only on the charges.
In particular, these supersymmetric equations show that the warp factor $U$ plays the role of a $c$-function for the flow.
The same attractive behaviour is exhibited by non-supersymmetric extremal black holes \cite{FK2,Goldstein:2005hq} and the warp factor is still playing the role of a $c$-function for these solutions \cite{Goldstein:2005rr}, but since they do not follow from (\ref{flows}), the construction of full solutions is significantly more difficult.

More recently, starting with \cite{CD}, it has become clear that non-BPS extremal black holes may enjoy the same properties of supersymmetric (BPS) ones, provided that one trades $|Z|$ with a different, real ``fake'' superpotential $W(z,\bar z )$ in (\ref{potential}) and (\ref{flows}). 
The extrema of 
\begin{equation}
	V_{BH}= W^2+4 g^{i{\bar\jmath}}	\partial_i W \partial_{\bar\jmath} W 
	\label{VBHW}
\end{equation}
and then of $W$ yield the non-supersymmetric attractor points as much as those of $Z$ describe the supersymmetric ones.
This procedure allows to find black hole solutions by solving first order differential equations for $W$
\begin{equation}
	U'=- e^U W\,,\qquad\  \qquad z'^i=-2 e^Ug^{i{\bar\jmath}}
	\partial_{\bar\jmath} W\,,  \label{nonBPSflows} 
\end{equation}
rather than the full second order equations of motion, even in absence of supersymmetry.
Moreover, $W$ is connected on the one hand to the value of the black hole entropy at the horizon $S_{bh} = \pi \,W^2$ and gives the ADM mass of the black hole ($M_{ADM} = W$) and the scalar charges $\Sigma_i = \partial_i W$ at infinity.
We stress that this formalism does not only allow to analyze the attractor point, like, for instance, the entropy function formalism \cite{entropyfunction}, but also the full attractor flow, from asymptotic infinity to the horizon.
As it obviously simplifies the task of building full solutions, this idea has been carefully analyzed in recent literature \cite{Andrianopoli:2007gt}--\cite{Andrianopoli:2009je}, producing new solutions and generalizations ranging from the $t^2$ to the $stu$ model, from static to rotating black holes, and also to extended supersymmetric and higher dimensional theories.

A central issue in this construction is whether $W$ always exists and what is its universal form.
In \cite{CD}, a general characterization of $W$ was given, together with a procedure to find at least one class of these ``superpotentials''. 
Moreover, $W$ was explicitly written in three simple instances: the $t^2$ single modulus (in full generality), the $t^3$ single modulus for restricted charge configurations and the $stu$ model for simple charge configurations and vanishing axions.
This class of solutions was later extended to the generating solutions of the $stu$ model by using a similar approach in 5-dimensions \cite{Cardoso:2007ky} (and the same solution, although in a different duality framework was later obtained in \cite{Hotta:2007wz,Gimon:2007mh} and constructed in its full form in \cite{Bellucci:2008sv}).
However, these examples made it clear that a deeper guideline was needed that better exploited the symmetries of the underlying theory. 

This guideline comes naturally from results in \cite{CDF} and \cite{FK1}. 
There, it was shown that the underlying special geometry of the scalar $\sigma$-model can be encoded in $Sp(2n+2)$ symplectic sections $(X^\Lambda,F_\Lambda)$ (where $\Lambda=0,\ldots,n)$ and the central charge is the symplectic product of these sections with the electric $q_\Lambda$ and magnetic charges $p^\Lambda$:  $Z=e^{K/2}(q_\Lambda X^\Lambda-p^\Lambda F_\Lambda)$. 
The symplectic sections also provide a projective parameterization of the scalar manifold, whose normal coordinates can be introduced by taking $t^i = X^i/X^0$.
Using these properties one can find the following differential identities for the central charge
\begin{eqnarray}
	D_i D_j Z &=& i C_{ijk}g^{k{\bar k}}{\overline {D}}_{\bar k}{\overline {Z}}\qquad (\overline{D}_{\bar\imath} C_{ijk}=0),\nonumber\\
	D_i D_{{\bar\jmath}}{\overline {Z}}&=& g_{i{\bar\jmath}}{\overline {Z}},\label{identities}\\
	D_i{\overline {Z}}&=&0\, , \nonumber 
\end{eqnarray}
and the curvature constraint 
\begin{equation}
	R_{i{\bar\jmath} k{\bar l}}=-g_{i{\bar\jmath}}g_{k{\bar l}}-g_{i{\bar l}}g_{k{\bar\jmath}}+C_{ikp}C_{{\bar\jmath}{\bar k}{\bar l}}g^{p{\bar p}}. \label{curvature} 
\end{equation}
The same structure also leads to the existence of some symplectic invariant quantities\footnote{When the scalar $\sigma$-model is described by a symmetric space $G/H$ this is equivalent to $H$-invariance. $H$-invariant quantities that extend to full $G$-invariant quantities are also moduli independent. An instance of a fully invariant quantity is the quartic invariant of symmetric special geometry $I_4(p^\Lambda, q_\Lambda)$.}, i.e.~quantities that do not change for a simultaneous symplectic action on the charge vector and on the scalar fields (defined through the symplectic sections $(X^\Lambda,F_\Lambda)$).
The simplest to identify are:
\begin{equation}
	I_1=|Z|^2 +|D_i Z|^2\qquad\qquad \hbox{and} \qquad\qquad I_2=|Z|^2-|D_i Z|^2\,.
\end{equation}
The black hole potential (\ref{potential}) coincides with one of them, namely $V_{BH}=I_1$, while the supersymmetric flow equations (\ref{flows}) are also driven by the invariant quantity $|Z| = \sqrt{(I_1 + I_2)/2}$. 
Then, it seems then quite natural to try and build also the ``fake'' superpotential $W$ in terms of symplectic invariants. 
This elegant approach was first attempted in \cite{Andrianopoli:2007gt} and gave some interesting results mainly for some of the extended supersymmetric theories, but still did not seem to capture the essence of $W$ for reasons that we will make evident below. 
Further progress on the form of $W$ was done in \cite{Ferrara:2008ap}, which revisited the expression for the quadratic series, and in \cite{Bellucci:2008sv}, where the general expression of $W$ for the $stu$ model was determined.
More recently, compelling evidence that the superpotential should have a definition in terms of symplectic invariant quantities has been provided in \cite{Andrianopoli:2009je}, where $W$ is identified with Hamilton's principal function associated to the non-BPS flow equations (\ref{nonBPSflows}).

The core question is: what is a complete set of duality invariant quantities? 
We can answer to this question by drawing a construction parallel to the $N=8$ case.
In $N=8$ supergravity there are 5 duality invariant quantities.
Four of them are given (in a polynomial way) by \cite{DAuria:1999fa} 
\begin{equation}
	{\rm tr} A, \quad 	{\rm tr} A^2, \quad	{\rm tr} A^3, \quad	{\rm tr} A^4, 
\end{equation}
where $A = Z Z^\dagger$ and $Z$ is the central charge matrix of the $N=8$ theory.
These invariant quantities are independent because they are related to the 4 eigenvalues of $ZZ^\dagger$.
There is a fifth independent invariant quantity that can be constructed in terms of the real part of the Pfaffian of $Z$:
\begin{equation}
	{\rm Re} ({\it Pf} \, Z)
\end{equation}
(note that $|{\it Pf}\, Z|^2$ is not independent because it can be expressed in terms of tr$A^4$, tr$A^3$tr$A$, (tr$A^2$)$^2$ and (tr$A$)$^4$).
The latter is the only invariant that is SU(8), but not U(8) invariant.
So there is a total of 5 invariant quantities and one relation among them, specified by the quartic Cartan invariant $I_4$.
A similar answer can be given for the $N=2$ case \cite{Cerchiai:2009pi}.
Besides
\begin{equation}
	i_1=Z{\overline {Z}}\qquad {\rm and} \qquad i_2= g^{i{\bar\jmath}} Z_i{\overline {Z}}_{\bar\jmath}\, , 
\label{i1i2}
\end{equation}
where $Z_i=D_iZ$, ${\overline {Z}}_{\bar\imath}={\overline {D}}_{\bar\imath} {\overline {Z}}$ and ${\overline Z}^i = g^{i \bar \jmath} \overline Z_{\bar \jmath}$,
three new invariants can be introduced: 
\begin{eqnarray}
	i_3&=&\frac16\left[ Z N_3({\overline {Z}})+{\overline {Z}} {\overline N}_3 (Z)\right], \label{i3}\\[3mm]	i_4 &=& \frac{i}{6}\left[ Z N_3({\overline {Z}}) - {\overline {Z}} {\overline N}_3(Z)\right]\,, \label{i4}\\[3mm]
	i_5 &=& g^{i{\bar\imath}}C_{ijk}C_{{\bar\imath}{\bar\jmath}\bar k}{\overline {Z}}^j{\overline {Z}}^k\, Z^{\bar\jmath} Z^{\bar k} \, , \label{i5}
\end{eqnarray}
where 
\begin{equation}
	N_3({\overline {Z}})=C_{ijk}{\overline {Z}}^i\ {\overline {Z}}^j\ {\overline {Z}}^k\,,\qquad \ \qquad {\overline N}_3(Z)=C_{{\bar\imath}{\bar\jmath}\bar k}Z^{\bar\imath}\ Z^{\bar\jmath}\ Z^{\bar k} .
\end{equation}
Also in this case there is one relation among them, which involves the quartic invariant $I_4$ of symmetric special geometry: 
\begin{equation}
	I_4=(i_1-i_2)^2+4 i_4-i_5\, . 
	\label{I4general}
\end{equation}
The crucial difference between the two cases is that the duality invariants are also $H$-invariants, and $H$ is completely different in $N=8$, where it is SU(8), and in $N=2$, where it is E$_6 \times$ U(1) (for the octonionic model).

Note that for symmetric spaces $\partial_i I_4=0$ follows from the additional properties
\begin{eqnarray}
&&D_k C_{ijk}= 0,\\
&&C_{j(lm} C_{pq)l} {\overline C}_{\bar\imath\bar\jmath{\bar k}}g^{j\bar\jmath}g^{k{\bar k}}=\frac43 C_{(lmp}g_{q){\bar l}}.
\end{eqnarray}

Based on the above considerations, the claim of this paper is that the ``fake" superpotential $W$ for the class of configurations corresponding to the non-BPS flows is given in terms of a non-polynomial expression of the purely duality invariant quantities $i_1$--$i_5$ (at least for symmetric special geometries). 
We will demonstrate this fact by first considering the quadratic and the cubic series and then illustrating the path towards the general case.

\section{The quadratic series}

Minimal couplings in special geometry \cite{Luciani} can be obtained through the quadratic series, based on holomorphic prepotentials of the form 
\begin{equation}
	F(X)=\frac{i}{2}\left[ (X^0)^2-\sum_{i=1}^{n}(X^i)^2\right]\, .
\end{equation}
The resulting moduli spaces are the coset manifolds $SU(1,n)/SU(n)\times U(1)$, which give simplified couplings in that they have a vanishing $C$-tensor: $C_{ijk}=0$.
This feature implies that non-BPS extremal black holes should have vanishing central charge at the horizon as follows from the attractor equations \cite{FGK}
\begin{equation}
	2 {\overline {Z}} D_i Z=-iC_{ijk}g^{j{\bar\jmath}}g^{k\bar k}{\overline {D}}_{\bar\jmath}{\overline {Z}}{\overline {D}}_{\bar k}{\overline {Z}} ,
\end{equation}
which, in this class of examples, reduce to 
\begin{equation}
	{\overline {Z}} D_i Z=0. \label{attractorquad}
\end{equation}
Since non-BPS fixed points correspond to $D_i Z\neq 0$, the central charge must vanish at the attractor point: $Z=0$.

According to our discussion in the introduction, full solutions for these non-BPS extremal black holes can be obtained from flow equations driven by a ``fake'' superpotential $W$.
The existence of such a function for the quadratic series of $N=2$ symmetric spaces, for arbitrary charge configurations and a single modulus, was established in \cite{CD} using the equivalent form of the prepotential $F(X)=-iX^0X^1$. 
The generalization to many moduli was then obtained in \cite{Andrianopoli:2007gt,Ferrara:2008ap}. 
We now show that the possibility of finding a simple expression of $W$ for these cases can be traced back to the fact that the superpotential must be constructed in terms of only two symplectic invariants:
\begin{equation}
	i_1=Z{\overline {Z}},\qquad\hbox{and}\qquad  i_2=g^{i{\bar\jmath}}D_i Z {\overline {D}}_{{\bar\jmath}} {\overline {Z}} .
\end{equation}
This happens because all the other invariants, namely $i_3$ (\ref{i3}), $i_4$ (\ref{i4}) and $i_5$ (\ref{i5}), depend on the $C_{ijk}$ intersection numbers and therefore they identically vanish for minimal couplings.
The moduli independent invariant, a quadratic form on charges, is $I_2=i_1-i_2$. 
Consistently the quartic invariant in (\ref{I4general}) becomes in this case $I_4=|i_1-i_2|^2 = I_2^2$.
$I_2$ is positive for BPS and negative for nonBPS attractor solutions.

As usual, the BPS flows are driven by $|Z| = \sqrt{i_1}$.
It turns out that in this case the non-BPS flows are driven by 
\begin{equation}
	W=\sqrt{i_2} = (g^{i{\bar\jmath}}D_i Z {\overline {D}}_{{\bar\jmath}} {\overline {Z}})^{1/2}\, . 
	\label{Wminimal}
\end{equation}
We will now prove this solely using the identities of special geometry (\ref{identities}) for the black hole central and matter charges, which simplify significantly when $C_{ijk}=0$ (so that $D_i D_j Z=0$).
In particular we will prove that the black hole potential can be expressed in terms of (\ref{Wminimal}), according to (\ref{VBHW}), and we will show that critical points of $W$ are in one-to-one correspondence with the non-BPS black hole configurations.
As a first step we obtain an explicit expression for the derivative of the superpotential by considering derivatives of $W^2$.
Because of the vanishing of the $C$-tensors these derivatives have a simple expression in terms of the central charge and its derivatives
\begin{equation}
	\partial_i \left(W^2\right) = 2 W	\partial_i W = D_i Z\, {\overline {Z}}\,,\qquad  \qquad 	\overline\partial_{\bar\imath} \left(W^2\right) = 2 W \overline\partial_{{\bar\imath}} W = {\overline {D}}_{{\bar\imath}} {\overline {Z}}\, Z\, ,
	\label{derWi}
\end{equation}
so that 
\begin{equation}
	\partial_i W =\frac{D_i Z{\overline {Z}}}{2(DZ {{\overline {D}} {\overline {Z}}}g^{-1})^{1/2}}\qquad \hbox{and} \qquad
	\overline \partial_{{\bar\imath}} W = \frac{{{\overline {D}}_{{\bar\imath}}{\overline {Z}}}Z}{2(DZ {{\overline {D}} {\overline {Z}}}g^{-1})^{1/2}}\,.
	\label{dWi}
\end{equation}
Then it is straightforward to show that 
\begin{equation}
	W^2=i_2=D_iZ {\overline {D}}_{{\bar\jmath}}{\overline {Z}} g^{i{\bar\jmath}},\qquad\qquad 4g^{i{\bar\jmath}}	\partial_iW	\partial_{\bar\jmath} W= i_1=i_2+I_2 = Z{\overline {Z}} ,
	\label{dWmin}
\end{equation}
so that the black hole potential can be written as
\begin{equation}
	V_{BH} = |Z|^2 + g^{i\bar\jmath}D_i Z \overline D_{\bar \jmath}\overline Z = W^2+4 g^{i{\bar\jmath}}	\partial_i W	\partial_{\bar\jmath} W\, . 
\end{equation}
In order to be a good superpotential, $W$ has to count the critical points of non-BPS black holes.
A simple inspection of (\ref{dWmin}) shows that this is indeed the case because at the non-BPS horizon $D_i Z$ must be non vanishing for some $i$ and thus
\begin{equation}
	\partial_i W=0 \quad \Leftrightarrow \quad Z=0 \,.
\end{equation}
Moreover, at the critical point $\partial_i W=0$ and therefore
\begin{equation}
	V_{BH} = W^2 = i_2 = -I_2,
\end{equation}
because $i_1=0$.
Note that for $n>1$ this model has $n-1$ complex flat directions, because (\ref{dWi}) vanishes in any direction once $Z = 0$.
Flat directions are a generic feature of all $Z \neq 0$ non-BPS models with the exception of the $t^3$ model \cite{Ferrara:2007tu}.

For the single modulus case ($n=1$) treated in \cite{CD}, it was found that the ``fake'' superpotential $W$ has an expression almost identical to the BPS superpotential, given by $|Z|$, up to some sign changes in the bare charges. 
This is now well understood because this model coincides with the bosonic sector of $N=4$ pure supergravity, where $Z$ and $D_t Z$ play the role of two eigenvalues of the central charge matrix and there is a complete symmetry between them.
The case with an arbitrary number of moduli is then a simple generalization of the one-modulus case, where, in the vein of \cite{CD}, the central charge and its derivative exchange their role.
In particular, it can be easily seen that the black hole potential described by the central charge and the one described by $W$ are the same because the gradients of ${\rm e}^U |Z|$ and ${\rm e}^U W$ in the enlarged moduli space (including also the warp factor) are related by a rotation matrix of the same form as the one of eq.~(4.10) in \cite{CD}.

\section{The cubic series} 

Another illustrative example is the simplest cubic model, having a single modulus, based on the prepotential
\begin{equation}
	F(X)=\frac{(X^1)^3}{X^0},
\end{equation} 
leading to the K\"ahler potential $K = -\log \left[-i(t-\bar t)^3\right]$.
The ``fake'' superpotential $W$ was obtained in \cite{CD} only for some very specific charge configurations and it was observed that the situation drastically changes with respect to the quadratic series.
The transformation from $Z$ to $W$ is field dependent and the ``fake'' superpotential is not simply related to the absolute value of the derivatives of the central charge. 
In the current framework this can be explained by the fact that there are now 3 non-zero independent duality invariants because $C_{ijk}\neq 0$.
These are $i_1$, $i_2$ and $i_3$, because, for a single modulus, we can prove that the other invariants are functionally dependent on them.
In detail, we find that
\begin{equation}
i_4 = -\sqrt{4\,\left(\frac{i_2}{3}\right)^3\,i_1-i_3^2}
\end{equation}
and
\begin{equation}
	i_5=\frac43 i_2^2,
\end{equation}
which means that we can also rewrite the expression for the quartic invariant (\ref{I4general}) as:
\begin{equation}
I_4=(i_1-i_2)^2-\frac43 i_2^2-4\sqrt{4\,\left(\frac{i_2}{3}\right)^3\,i_1-i_3^2}\,.
\end{equation}
The main new result we present in this section is the explicit expression of $W$ for all charges and given in terms of the symplectic invariants mentioned above.

Since we are going to prove that $W$ can be given only in terms of symplectic invariant quantities, we can compute it in any frame, i.e.~for any charge configuration, and only later check that we did not miss any term by switching on all the charges in the final expression.
Therefore we start from a D0-D6 configuration, involving only $q_0$ and $p^0$.
When all the other charges are vanishing, the only allowed black hole configurations are non-BPS.
Hence this case is a natural representative of the non-BPS branch.
The form of the superpotential for this case can be obtained by properly identifying all the moduli in the analogous setup in the context of the $stu$ model \cite{Bellucci:2008sv}.
The resulting expression is 
\begin{equation}
	W^2={\rm e}^K\left|(q_0)^{1/3}+(p^0)^{1/3} t\right|^2  \left[(q_0)^{2/3}-\frac{1}{2}(p^0 q_0)^{1/3} (t+\overline{t})+ (p^0)^{2/3} t \overline{t}\right]^2 
	\label{Wt3}
\end{equation}
and we have checked that it fulfills (\ref{VBHW}).

We are now going to rewrite (\ref{Wt3}) in terms of the independent symplectic invariants, which, for this choice of charges, read
\begin{eqnarray}
	 i_1&=& {\rm e}^K \left|q_0+p^0 t^3\right|^2 ,\\[2mm]
	 i_2&=& 3{\rm e}^K \left[(q_0)^2+q_0p^0 t\overline{t}(t+\overline{t})+(p^0)^2 t^3 \overline{t}^3\right],\\[2mm]
	 i_3&=& {\rm e}^K \sqrt{-I_4} \left[(q_0)^2-(p^0)^2 t^3 \overline{t}^3\right],
\end{eqnarray}
where $I_4 = - \left(p^0q_0\right)^2$.
We start by computing the difference between the square of the ``fake'' superpotential and the square of the absolute value of the central charge:
\begin{equation}
	W^2-i_1=- \frac34 {\rm e}^K \sqrt{-I_4} (t-\bar t)^2  \left[(q_0/p^0)^{1/3}+(p^0/q_0)^{1/3}t\overline{t}+(t+\overline{t})\right]. 
	\label{l0}
\end{equation}
Then we need to rewrite the three real terms in the square brackets above as combinations of the symplectic invariants. 
The easiest one to identify is the last term, which is proportional to the following difference:
\begin{equation}
	i_1-\frac{i_2}{3}={\rm e}^K \sqrt{-I_4}(t-\overline{t})^2 (t+\overline{t}).
	\label{l1} 
\end{equation}
Upon using this relation in (\ref{l0}) we obtain that
\begin{equation}
	W^2=\frac{i_1+i_2}{4}-\frac34 {\rm e}^K  \sqrt{-I_4}(t-\overline{t})^2 \left[(q_0/p^0)^{1/3}+(p^0/q_0)^{1/3}t\overline{t}\right]. 
\end{equation}
The two remaining terms can also be expressed in terms of invariants, though with some effort.
First we can get rid of the $q_0 p^0$ combination in $i_2$ by considering the following expression 
\begin{equation}
	\frac{i_1+i_2}{4}-\frac{1}{4I_4}\left(i_1-\frac{i_2}{3}\right)^3={\rm e}^K  \left[(q_0)^2+(p^0)^2 t^3 \overline{t}^3\right].
\end{equation}
Then, using this expression and the one of $i_3$ we can eventually construct the necessary combinations 
\begin{eqnarray}
-8 I_4\,	{\rm e}^K q_0^2= \left(i_1-\frac{i_2}{3}\right)^3- I_4(i_1+i_2)+4 i_3 \sqrt{-I_4} \,,\\[3mm]
-8I_4\,	{\rm e}^K (p^0)^2\,  t^3\overline{t}^3=\left(i_1-\frac{i_2}{3}\right)^3- I_4(i_1+i_2)-4 i_3 \sqrt{-I_4} \,.
\end{eqnarray}
Putting all these ingredients together we reach the final expression for the ``fake'' superpotential, which reads:
\begin{equation}
	\begin{array}{rcl}
		W^2&=&\displaystyle \frac{i_1+i_2}{4}+\frac{3}{8}\left[\left(\left(i_1-\frac{i_2}{3}\right)^3-(i_1+i_2)\,I_4+4\, i_3\,\sqrt{-I_4}\right)^{1/3}+ \right. \\[5mm]
		&&\displaystyle \left.+\left(\left(i_1-\frac{i_2}{3}\right)^3-(i_1+i_2)\,I_4- 4\, i_3\,\sqrt{-I_4}\right)^{1/3}\right].
	\end{array}
	\label{geom} 
\end{equation}
We can now see that in this case the superpotential is not simply given by a linear combination of the invariants, but rather by a non-polynomial expression (containing radicals) of the basic duality invariant quantities.
At the attractor point we have that
\begin{equation}
	i_2 = 3 i_1 = \frac34 \sqrt{-I_4}
\end{equation}
and $i_3$ vanishes.
Hence the above expression reduces to 
\begin{equation}
	W^2 = \sqrt{-I_4},
\end{equation}
which is the expected result for a non-BPS black hole.

Although we have computed the ``fake'' superpotential in terms of the symplectic invariants in the special case of vanishing $q_1$ and $p^1$ charges, the final expression should be the same in any duality frame. 
This can be checked by comparing the expression for $W^2$ resulting by evaluating (\ref{geom}) with all the charges  $(q_0,\,q_1,\,p^0,\,p^1)$  switched on and the analogous one coming from the $stu$ model discussed in \cite{Bellucci:2008sv}, in the $S=T=U=t$ limit.
As expected, the two expressions agree.

We report here for completeness the explicit form of the superpotential with all the charges:
\begin{eqnarray}
	\nonumber W^2=&&{\rm e}^K \frac{\sqrt{-I_4} \left(-1+\nu ^3\right)^2}{\nu^3 (\sigma_- +\sigma_+ )^3}\left(t \overline{t}+(t+\overline{t}) \frac{\sigma_- +\nu \sigma_+ }{ 1-\nu }+\left(\frac{\sigma_- +\nu \sigma_+ }{ 1-\nu }\right)^2\right) \\
	&&\left(t \overline{t} +(t+\overline{t}) \frac{((2+\nu ) \sigma_- -\nu (1+2 \nu ) \sigma_+ )}{2 \left(1+\nu +\nu ^2\right)}+\frac{\sigma_- ^2-\nu \sigma_- \sigma_+ +\nu ^2 \sigma_+ ^2}{1+\nu +\nu ^2}\right)^2\label{wquad} 
\end{eqnarray}
where 
\begin{eqnarray}
	&&\nu =\left(\frac{2 (p^1)^3+p^0 \left(\sqrt{-I_4}-p^0 q_0-p^1 q_1\right)}{2 (p^1)^3-p^0 \left(\sqrt{-I_4}+p^0 q_0+p^1 q_1\right)}\right)^{1/3},\label{defnu}\\[3mm]
	&&\nonumber \sigma_\pm =\frac{1}{2} \frac{\sqrt{-I_4}\pm p^0 q_0\pm \frac{p^1 q_1}{3}}{(p^1)^2-\frac{p^0 q_1}{3}}. \label{defrhosigma}
\end{eqnarray}

\section{Full Non-BPS solution}

Given the most general form of the superpotential for the single modulus cubic model, we can now solve the flow equations (\ref{nonBPSflows}) and provide an explicit form for the most general non-BPS black hole configuration in this context.

\begin{figure}[ht]
	\centering
		\includegraphics[scale=.83]{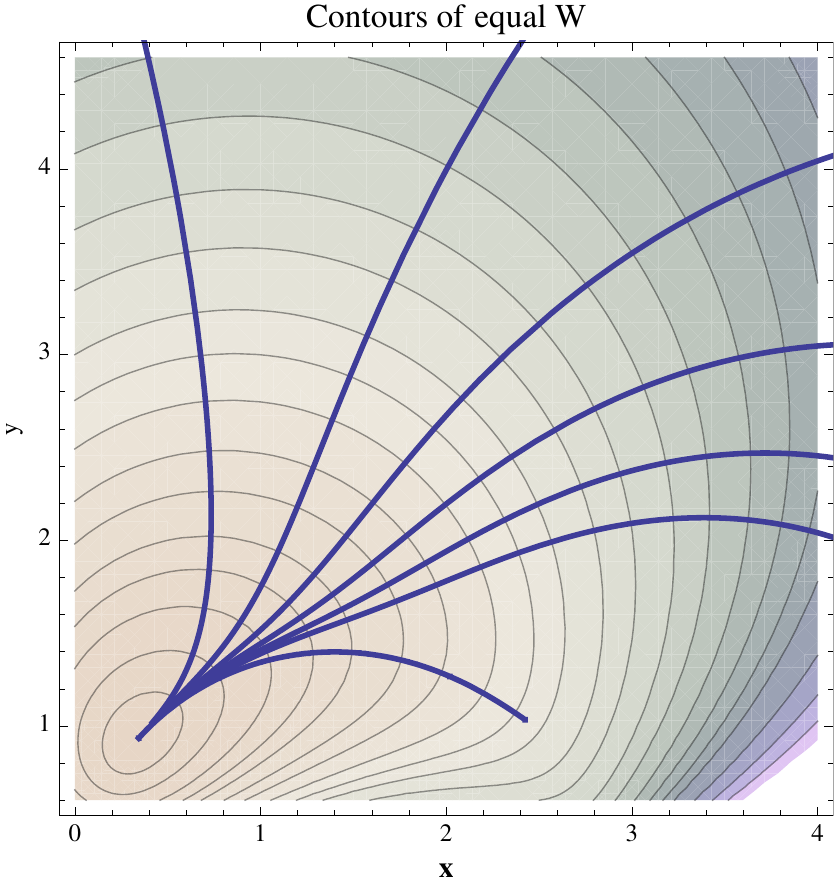}\qquad
		\includegraphics[scale=.83]{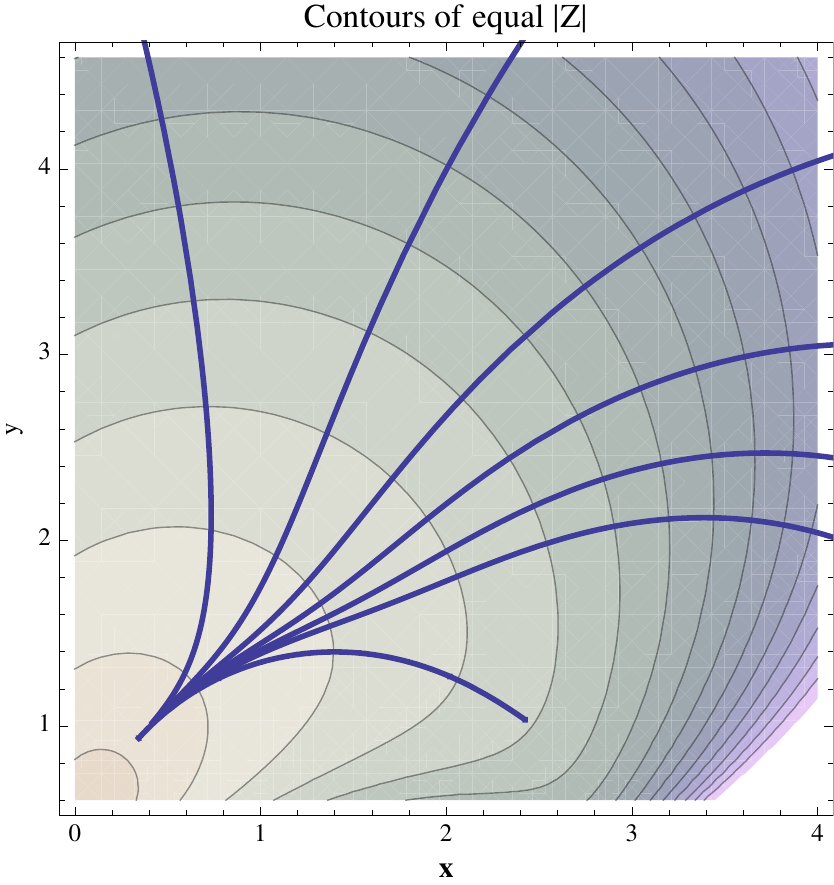}
\caption{These contour plots provide the values of $W$ and $|Z|$ in the $x,y$ moduli space for a generic non-BPS configuration with charges $p^1=p^0=q_1=1$, $q_0=-1$. 
The thick blue lines correspond to gradient flows of the scalar fields according to (\ref{solgeneral}) for different asymptotic values of the scalars. 
It is clear that while these flows tend towards a critical point of $W$, the attractor point is at a generic value of $|Z|$.}
	\label{figura}
\end{figure}

\begin{figure}[hpbt]
	\centering
		\includegraphics[scale=.83]{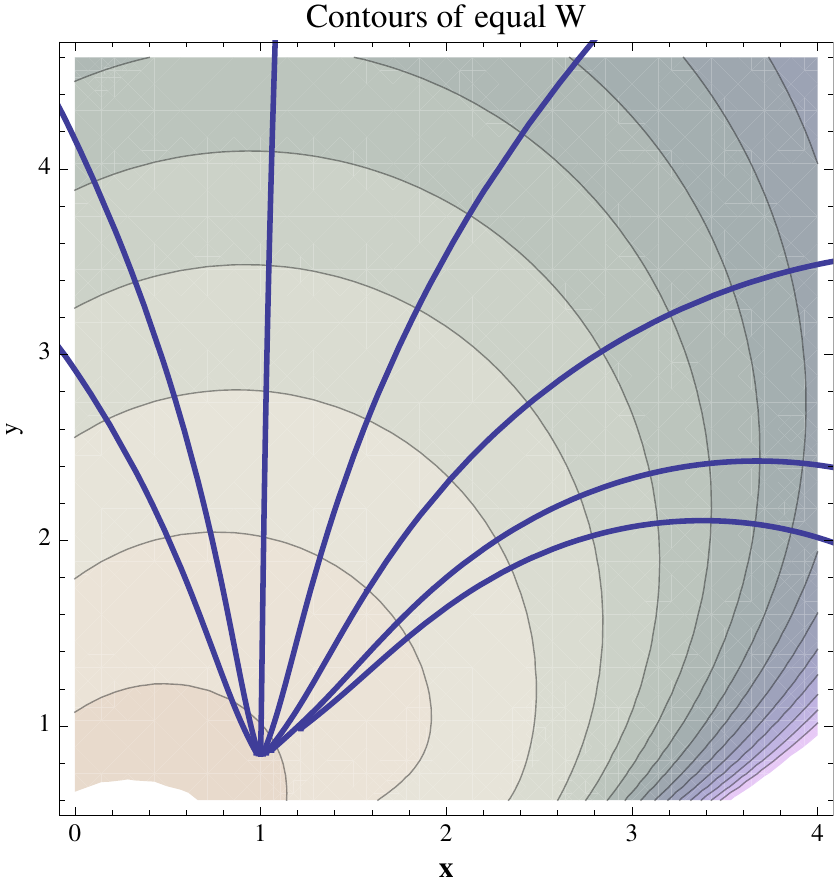}\qquad
		\includegraphics[scale=.83]{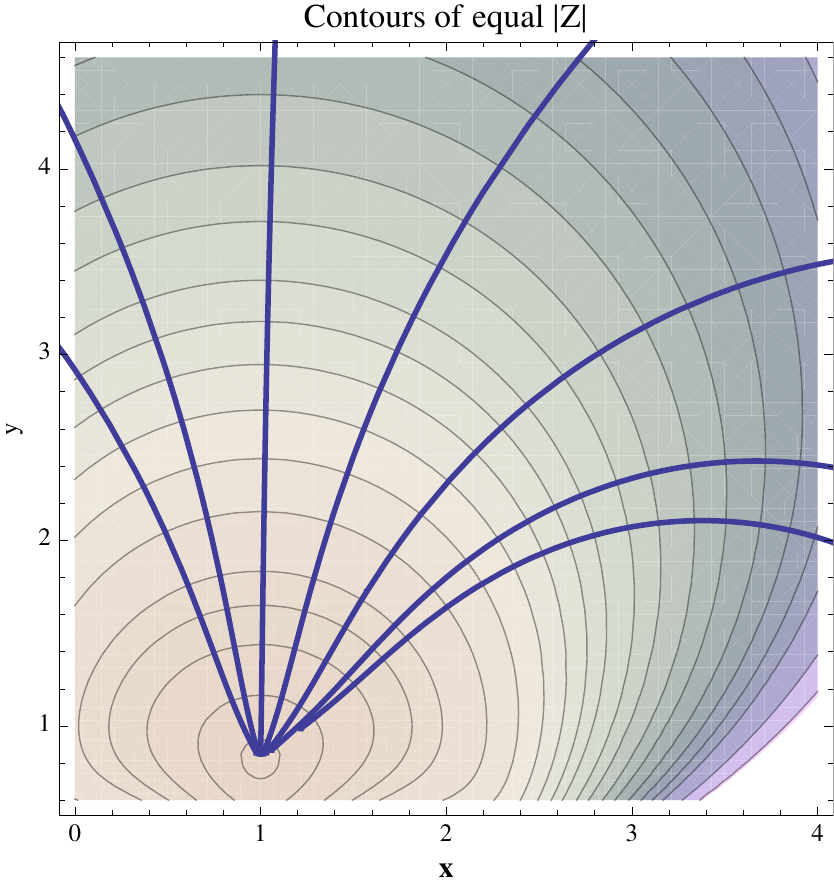}
\caption{These contour plots provide the values of $W$ and $|Z|$ in the $x,y$ moduli space for a generic supersymmetric configuration with charges $p^1=p^0=q_1=q_0=1$. 
The thick blue lines correspond to gradient flows of the scalar fields according to (\ref{flows}) for different asymptotic values of the scalars. 
In this case the flows approach a critical point of the central charge $|Z|$, which has no special meaning in terms of $W$.}
	\label{figura2}
\end{figure}

Again, it is easier to present first the computation of the solution in a specific duality frame and then boost this solution to the most general one containing all the charges, although one could equally start directly from (\ref{wquad}) and solve the corresponding equations.
If we set $p^0=q_1=0$ in (\ref{geom}) and assume $q_0<0$ and $p^1>0$, the superpotential reduces to a rather simple expression:
\begin{eqnarray}
	W= {\rm e}^{K/2} (-q_0 + 3 p^1 t \bar t)=\frac{-q_0+3 p^1 \left(x^2+y^2\right)}{2 \sqrt{2} \sqrt{y^3}},
	\label{superpotentialq0p1} 
\end{eqnarray}
where we have replaced the complex scalar field by its real and imaginary components following $t = x - i y$.
Introducing the definition 
\begin{equation}
	a \equiv {\rm e}^{-U},
	\label{def}
\end{equation}
we can rewrite the flow equations (\ref{nonBPSflows}) for the superpotential (\ref{superpotentialq0p1}) as the following coupled system of equations
\begin{eqnarray}
a^\prime & = & W(x,y,q_0,p^1), \label{flow1} \\[2mm]
a x^\prime & = & -2 \sqrt{2} p^1 x \sqrt{y}, \label{flowx} \\[2mm]
a y^\prime &=& -\frac{q_0 + p^1 (-3 x^2 + y^2)}{\sqrt{2y}}.\label{flowy}
\end{eqnarray}
The solution to this system can be obtained in terms of harmonic functions by taking appropriate combinations of the fields.
First, we can see that the right hand side of (\ref{flowy}) is very close to $W$ (which is $a'$, according to (\ref{flow1})) and hence we can get rid of $q_0$ by taking the combination
\begin{equation}
	\left(\frac{a}{\sqrt{y}}\right)^\prime = \sqrt{2}p^1.
	\label{combo1}
\end{equation}
Then we can solve this equation by introducing the harmonic function
\begin{equation}
	H^1 = h^1 + \sqrt{2} p^1 \tau,
	\label{H1}
\end{equation}
so that
\begin{equation}
	\frac{a}{\sqrt{y}} = H^1.
	\label{sol1}
\end{equation}
Then we can solve for $x$, by replacing both the combination $a/\sqrt{y}$ and $p^1$ in (\ref{flowx}) by $H^1$ and its derivative, so that
\begin{equation}
	\frac{x'}{x} = - 2 \frac{\left(H^1\right)^\prime}{H^1},
\end{equation}
whose solution reads
\begin{equation}
	x = \frac{b}{(H^1)^2} .
	\label{xsol}
\end{equation}
Finally we can take another combination of the warp factor and $y$ that gets rid of $p^1$, upon using (\ref{xsol}):
\begin{equation}
		\left(a y^{3/2}\right)^\prime = \sqrt{2}(-q_0 + 3 p^1 x^2) = -\sqrt{2} q_0 + 3 b^2 \frac{(H^1)^\prime}{(H^1)^4}. 
	\label{combo2}
\end{equation}
This can be easily integrated by introducing another harmonic function
\begin{equation}
	H_0 = h_0 - \sqrt{2}\, q_0\, \tau
\end{equation}
and gives
\begin{equation}
	a y^{3/2} = H_0 - \frac{b^2}{(H^1)^3}.
	\label{sol2}
\end{equation}
We can then put together (\ref{sol1}) and (\ref{sol2}) with the appropriate powers to get an expression for the warp factor and dilaton field as:
\begin{equation}
	a^4 = (H^1)^3 H_0 - b^2, \qquad y = \frac{a^2}{H_1^2}.
	\label{final1}
\end{equation}
We can also rewrite this solution in terms of the quartic invariant $I_4 = 4 (p^1)^3 q_0$ and new, rescaled, integration constants 
\begin{equation}
	b_0 = \frac{h_0}{\sqrt{2}} \frac{(-I_4)^{1/4}}{q_0}, \qquad	b^1 = -\frac{h^1}{\sqrt{2}} \frac{(-I_4)^{1/4}}{p^1}, 
\end{equation} 
by introducing ``universal'' harmonic functions:
\begin{eqnarray}
  {\cal H}_0 &=& (b_0 - (-I_4)^{1/4} \tau) = \frac{(-I_4)^{1/4}}{\sqrt{2}q_0} H_0, \\[2mm]
  {\cal H}_1 &=& (b_1 - (-I_4)^{1/4} \tau) = -\frac{(-I_4)^{1/4}}{\sqrt{2}p^1} H^1.
\end{eqnarray}
The solution then reads
\begin{eqnarray}
	{\rm e}^{-4U} &=& ({\cal H}_1)^3 {\cal H}_0 - b^2, \nonumber \\[2mm]
	x &=& \frac{b\sqrt{-I_4}}{2 (p^1)^2({\cal H}_1)^2}, \label{solspecial} \\[2mm]
	y &=& \frac{{\rm e}^{-2 U}\sqrt{-I_4}}{2 (p^1)^2 {\cal H}_1^2}.\nonumber
\end{eqnarray}

The ``universal'' harmonic functions are better suited to obtain the general solution starting from the seed solution constructed above, because they are built in terms of the quartic invariant $I_4$.
After using the transformation technique we generate the most general solution with all charges switched on. 
This solution can be expressed in a compact form by using the charge combinations $\nu$ and $\sigma_\pm$, defined in (\ref{defnu}) and (\ref{defrhosigma}),
\begin{eqnarray}
{\rm e}^{-4 U}&=&({\cal H}_1)^3 {\cal H}_0 - b^2,\nonumber\\[3mm]
	x&=&\frac{{\cal H}_1^2(\nu+1)(\nu \sigma_+-\sigma_-)+{\cal H}_0{\cal H}_1 (\nu \sigma_++\sigma_-)(\nu-1)+2b(\sigma_+\nu^2+\sigma_-)}{(\nu+1)^2{\cal H}_1^2+(\nu-1)^2{\cal H}_0{\cal H}_1+2b(\nu^2-1)}, \label{solgeneral}\\[3mm]
	y&=&\frac{2\nu(\sigma_-+\sigma_+){\rm e}^{-2 U}}{(\nu+1)^2{\cal H}_1^2+(\nu-1)^2{\cal H}_0{\cal H}_1+2b(\nu^2-1)}, \nonumber
\end{eqnarray}
and it correctly reduces to (\ref{solspecial}) when $p^0 = q_1 =0$, which implies that $\nu = 1$ and $\sigma_\pm = \frac12 \frac{\sqrt{-I_4}}{(p^1)^2}$.
The attractor values of the above solution are:
\begin{eqnarray}
{\rm e}^{-2 U(\tau\to - \infty)}&=&\sqrt{-I_4}\tau^2,\nonumber\\[3mm]
	x(\tau\to - \infty)&=&\frac{\sigma_+ \nu^2- \sigma_-}{1+ \nu^2}, \label{generattract}\\[3mm]
	y(\tau\to - \infty)&=&\frac{(\sigma_+ + \sigma_-)\nu}{1+ \nu^2}. \nonumber
\end{eqnarray}

\section{General case and outlook} 

It is obvious that the results for the cubic series should have a simple generalization to the case of an arbitrary number of moduli, at least for all special geometries based on symmetric spaces. 
However, we note that, when we have more than one modulus, $i_5$ becomes an independent invariant. 
For example, in the $stu$ model 
\begin{equation}
	i_5=|D_s Z|^2 |D_t Z|^2+|D_s Z|^2 |D_u Z|^2 +|D_t Z|^2 |D_u Z|^2\, , 
\end{equation}
which cannot be written in terms of the other invariants. 
Actually, $I_4$ depends on $i_5$ and so should the superpotential $W$. 
Therefore we expect that for a generic model
\begin{equation}
	W=W(i_1, i_2, i_3, i_4, i_5) .
\end{equation}
It is also clear that previous attempts to find $W$ failed either because a too na\"ive generalization of the quadratic case was proposed, with $W$ as a simple linear combination of $i_1$ and $i_2$, or because the other invariants, namely $i_3$, $i_4$ and $i_5$, were not included.

We also remark that non-BPS solutions with $Z=0$ should have a simple superpotential in virtue of the fact that they can be embedded in BPS solutions coming from extended ($N>2$) supergravities \cite{Andrianopoli:2007gt,Ferrara:2008ap}. 
For the quadratic series, non-BPS solutions with $Z=0$ are the only possible ones, while for cubic geometries, these solutions start to arise when two or more moduli are present ($F=st^2$ and $F=stu$). 
For instance, in the $stu$ model a $Z=0$ superpotential can be obtained by interchanging $|Z|$ with $|D_sZ|$ so that 
\begin{equation}
	W=|D_sZ|.
\end{equation}
This result was already obtained in \cite{Bellucci:2008sv} in a specific symplectic frame but here we exhibit its derivation in a duality invariant formulation, valid in any frame.

Using the factorization of the $stu$ moduli space in three (complex) one-dimensional manifolds, from (\ref{curvature}) one gets the following interesting relations between the curvature, the $C$-tensors and the metric of the scalar $\sigma$-model:
\begin{eqnarray}
R_{i\bar{\imath} i\bar{\imath}} &=& -2g_{i\bar{\imath}}g_{i\bar{\imath}},\qquad i=s,t,u,\\
C_{stu}{\overline C}_{{\bar s}{\bar t}{\bar u}}&=& g_{s\bar s}g_{t\bar t}g_{u\bar u}\,.
\end{eqnarray}
This means that, by choosing the phase such that $C_{stu}$ is real, we can define $C_{stu}=g_{s {\bar s}}^{1/2}g_{t {\bar t}}^{1/2}g_{u {\bar u}}^{1/2}$ so that the special geometry identities (\ref{identities}) become
\begin{equation}
D_s D_t\,  Z=i g_{s {\bar s}}^{1/2}g_{t {\bar t}}^{1/2}g_{u {\bar u}}^{-1/2} \overline D_{\bar u} \overline{Z} \qquad (\hbox{and the same for}\ s\to t\to u)\, .
\end{equation}
Factorization of the moduli space also implies that one can write three separate $i_2$ invariants, one per each modulus $i_2^s = |D_sZ|^2$, $i_2^t = |D_tZ|^2$ and $i_2^u = |D_uZ|^2$.
Then it is easy to see that a good invariant superpotential is
\begin{equation}
W=\sqrt{i_2^s} = |D_sZ D_{\bar s} \overline{Z}\, g^{s{\bar s}}|^{1/2}=|D_s\, Z|\, .
\end{equation}
Indeed, by differentiating $W^2$ we find
\begin{eqnarray}
2\partial_{\bar t}W\, W  &=&-i g_{t {\bar t}}^{1/2}g_{u {\bar u}}^{-1/2}g_{s {\bar s}}^{-1/2} D_u Z D_s Z,\\[2mm]
2\partial_{\bar u}W\, W &=&-i g_{u {\bar u}}^{1/2}g_{t {\bar t}}^{-1/2}g_{s {\bar s}}^{-1/2} D_t Z D_s Z,\\[2mm]
2\partial_{\bar s}W\, W &=& Z D_s Z,
\end{eqnarray}
and therefore
\begin{eqnarray}
4g^{t{\bar t}}\partial_{\bar t} W \partial_t W&=&i_2^u = |D_u Z|^2,\\
4g^{u{\bar u}}\partial_{\bar u} W \partial_u W&=&i_2^t = |D_t Z|^2,\\
4g^{s{\bar s}}\partial_{\bar s} W \partial_s W&=& i_1 = | Z|^2\, .
\end{eqnarray}
We have just completed the proof because the previous relations show that the superpotential verifies
\begin{eqnarray}
&&W^2 = |D_s Z|^2,\\
&&4(g^{s{\bar s}}\partial_{\bar s} W \partial_s W+g^{t{\bar t}}\partial_{\bar t} W \partial_t W+g^{u{\bar u}}\partial_{\bar u} W \partial_u W)=|Z|^2+|D_u Z|^2+|D_t Z|^2\,,
\end{eqnarray}
as it should be, and the attractor points occur at $D_s Z\neq 0$, $Z=D_u Z=D_t Z=0$, which is equivalent to the request that $\partial_s W = \partial_t W = \partial_u W =0$. 

We conclude by pointing out that the previous derivation admits a straightforward extension to all the models in the series SU(1,1)/U(1) $\times$ SO($2,2+n$)/SO(2)$\times$SO($2+n$).
Any model in this class admits non-BPS black holes with $Z = 0$ whose superpotential can be identified with $W = |D_s Z|$, where $s$ is the modulus of the SU(1,1)/U(1) factor.
This follows again from the factorization of the scalar manifold, which implies that
\begin{equation}
	R_{s \bar s i \bar \imath} = 0 \quad \Longrightarrow \quad C_{sik} C_{\bar s\bar \imath \bar k} g^{k \bar k} = g_{s \bar s} g_{i \bar \imath}.
\end{equation}
This can be used to prove that
\begin{equation}
	4 |\partial_i W|^2 = |D_i Z|^2 \quad \hbox{and}	\quad 4 |\partial_s W|^2 = |Z|^2,
\end{equation}
which, together with the definition of the superpotential $W^2 = |D_s Z|^2$, gives the right potential and non-BPS critical points.
Note, however, that for $n>1$ other $Z = 0$ non-BPS solutions exist with flat directions for which the ``fake superpotential'' $W$ cannot be constructed in this simple way \cite{Bellucci:2006xz}.

\begin{table}[htb]
	\begin{center}
		\begin{tabular}{cccccc}\hline\\[-3mm]
		&quadratic   &  & $stu$  && $t^3$ model\\
		&series  &  & model && \\
		& $Z=0$ && $Z=0$ && $Z\neq 0$\\[-3mm]\\\hline \\
		$W^2$ &$i_2$ && $i_2^s$&& $\begin{array}{l}
		(i_1+i_2)/4+\\[2mm]+\nicefrac{3}{8}[(4\, i_3\,\sqrt{-I_4}-(i_1+i_2)\,I_4+(i_1-\nicefrac{i_2}{3})^3)^{1/3}+\\[2mm]
		+ (-4\, i_3\,\sqrt{-I_4}-(i_1+i_2)\,I_4+(i_1-\nicefrac{i_2}{3})^3)^{1/3}]
		\end{array}$ \\ 
		\\
		$4 g^{i\bar \jmath} \partial_i W \partial_{\bar\jmath}W$ & $i_1 = i_2 + I_2$&& $i_1+i_2^t+i_2^u$&& $\begin{array}{l}
		\nicefrac{3}{4}(i_1+i_2)-\\[2mm]-\nicefrac{3}{8}[(4\, i_3\,\sqrt{-I_4}-(i_1+i_2)\,I_4+(i_1-\nicefrac{i_2}{3})^3)^{1/3}+\\[2mm]
		+ (-4\, i_3\,\sqrt{-I_4}-(i_1+i_2)\,I_4+(i_1-\nicefrac{i_2}{3})^3)^{1/3}]
		\end{array}$\\
		\\\hline
		\end{tabular}
	\end{center}
Summary Table: \emph{Superpotential and derivatives of the superpotential in terms of invariants for the models considered in this letter. While the $Z=0$ cases have simple polynomial expressions, the $Z \neq 0$ case is significantly more complicated.}
\end{table}

\bigskip
\section*{Acknowledgments}

\noindent We gladly acknowledge the stimulating environment offered by the STRINGS 2009 conference in Rome and the SAM school 2009 in Frascati.
This work is supported in part by the ERC Advanced Grant no. 226455, \textit{``Supersymmetry, Quantum Gravity and Gauge Fields''} (\textit{SUPERFIELDS}). 
The work of A.~C.~is partially supported by MIUR-PRIN contract 20075ATT78, the work of G.~D.~has been partially supported by the Fondazione Cariparo Excellence Grant {\em String-derived supergravities with branes and fluxes and their phenomenological implications}, the work of S.~F.~has been supported in part by D.O.E.~grant DE-FG03-91ER40662, Task C and the work of A.~Y.~has been supported in part by CERN-PH-TH where part of the work was done.


\end{document}